\documentclass[12pt]{article}
\usepackage{amssymb}
\begin{document}\def\ov{\over} \def\ep{\varepsilon}
\newcommand{\C}[1]{{\cal C}_{#1}} \def\inv{^{-1}}\def\be{\begin{equation}}
\def\ee{\end{equation}}\def\x{\xi}\def\({\left(} \def\){\right)} \def\iy{\infty}
 \def\e{\eta} \def\cd{\cdots} \def\ph{\varphi} \def\ps{\psi} \def\La{\Lambda} \def\s{\sigma} \def\ds{\displaystyle}
 \newcommand{\sst}{\scriptstyle}
\newcommand{\xs}[1]{\x_{\s(#1)}} \newcommand\sg[1]{{\rm sgn}\,#1}
\newcommand{\xii}[1]{\x_{i_{#1}}} \def\ld{\ldots} \def\Z{\mathbb Z}
\def\a{\alpha}\def\g{\gamma}\def\b{\beta} \def\noi{\noindent}
\def\ar{\leftrightarrow} \def\d{\delta}
 \def\S{\mathbb S} \def\tl{\tilde}
  \def\E{\mathbb E} \def\P{\mathbb P}
  \newcommand{\R}{\mathbb{R}}
  \newcommand{\pl}{\partial}
  \newcommand{\sq}{\sqrt}
  \newcommand{\ra}{\rightarrow}
  \newcommand{\CR}{\mathcal{R}}
 \def\CS{{\mathcal S}}
  \newcommand{\br}[2]{\left[{#1\atop #2}\right]}
\def\bs{\backslash}\def\N{\mathbb N}
\newcommand{\ba}{\begin{eqnarray*}}
\newcommand{\ea}{\end{eqnarray*}}
\hfill August 18, 2008
\begin{center}{\Large\bf The Dynamics of the One-Dimensional\vspace{1ex}\\
Delta-Function Bose Gas}
\end{center}

\begin{center}{\large\bf Craig A.~Tracy}\\
{\it Department of Mathematics \\
University of California\\
Davis, CA 95616, USA\\
email:\texttt{tracy@math.ucdavis.edu}}\end{center}

\begin{center}{\large \bf Harold Widom}\\
{\it Department of Mathematics\\
University of California\\
Santa Cruz, CA 95064, USA\\
email: \texttt{widom@ucsc.edu}}\end{center}

\begin{abstract}
We give a method to
 solve the time-dependent Schr\"odinger equation for a system of
  one-dimensional bosons interacting via a repulsive delta function potential.  
  The method uses the ideas of Bethe Ansatz but does not use the spectral theory
  of the associated Hamiltonian.
\end{abstract}
\newpage

\section{Introduction}
In this paper we give an alternative approach to solve the time-dependent
Schr\"odinger equation for a system of one-dimensional bosons interacting
via a delta function potential.  This quantum many-body model, called the 
\textit{Lieb-Liniger model} \cite{L, LL},  is   
``exactly solvable'' by  means of Bethe Ansatz \cite{Be} in the sense that the ground state energy \cite{LL}, the excitation spectrum \cite{L} as well as equilibrium thermodynamics  \cite{YY} are  known.  (See \cite{KBI, Su2}  for textbook treatments.)
Current widespread  interest  in  the Lieb-Liniger model has arisen because of  its connection to
  ultracold gases confined in a quasi one-dimensional trap. 
Indeed, it has been recently shown that the Lieb-Liniger model for one-dimensional bosons ``can be rigorously
derived via a scaling limit from a dilute three-dimensional Bose gas with arbitrary repulsive interaction potential
of finite scattering length'' \cite{Se}.   These applications focus interest
on the time-dependent solutions.  For a   review of these developments
 see \cite{BDZ}.

Recall that the Lieb-Liniger  $\d$-function ($N$-particle) Bose gas Hamiltonian  is
\be H=-\sum_{j=1}^N {\pl^2\ov \pl x_j^2} + 2c \sum_{ j<k} \d(x_j-x_k) \label{H}\ee
where $c > 0$ (repulsive case) and 
$x_j\in\R$.\footnote{We choose units where $2m=1$
and $\hbar=1$.}
Most  work \cite{BGOL, Ga, L, LL, Su1, Ya} focuses on the eigenfunctions and eigenvalues  of $H$ from which,
in principle, any  solution $\Psi(x;t)$ of the time-dependent Schr\"odinger equation 
\be H\Psi = i\,{\pl \Psi\ov \pl t}, \label{Schr} \ee
subject to the  initial condition 
\be \Psi(x;0)=\Psi(x_1,\ldots, x_N;0)= \psi_0(x_1,\ldots, x_N), \label{IC} \ee
 may be constructed.\footnote{$\psi_0$ is a given symmetric function of the coordinates $x_j$.}
In practice, due to the complexity of the
Bethe eigenfunctions and the nonlinear Bethe equations, it is difficult to analyze time-dependent solutions whose initial
conditions are not eigenfunctions.  We give here an alternative to the spectral method for solving (\ref{Schr}).

Since
different physical conditions require different choices of the initial wave function $\psi_0$,
we develop here a flexible  mathematical structure  to incorporate
these different choices.
Suppose we solve (\ref{Schr}) with the initial condition
\be \psi_\d(x)=\psi_\d(x_1,\ldots,x_N)={1\ov N!}\,\sum_{\s\in\S_N}\prod_{j=1}^N \d\left(x_{\s(j)}-y_j\right)
\label{deltaInitial}\ee
where $\S_N$ is the permutation group acting on $\{1, \ldots, N\}$.
Here $y_j\in\R$ are fixed and without loss of generality we may  assume
\be y_1<y_2<\cdots < y_N.\label{yOrder}\ee
If $\Psi_\d(x,y;t)$ denotes this solution, 
then the solution to (\ref{Schr}) satisfying (\ref{IC}) is
\[ \Psi(x;t)=\int_{\R^N} \Psi_\d(x,y;t)\psi_0(y) \, dy \]
($dy=dy_1\cdots dy_N$).  The subject  of this paper is $\Psi_\d$, a Green's function for (\ref{Schr}).

We remark that there are two natural choices for the initial wave function $\psi_0$.
The free particle Gaussian wave function localized at $y$ 
	is\footnote{Thus $\left|\ph_a(x;y,p)\right|^2$ is the Gaussian distribution centered at $y$ with variance $a^2$.  Solving the time-dependent \textit{free} Schr\"odinger equation shows that $\langle x \rangle=2pt$; and hence, $2p$ is the classical velocity. (Recall $2m=1$.)}
\[ \ph_a(x;y,p)= {1\ov (2\pi)^{1/4} \sq{a}}\, e^{-(x-y)^2/(4a^2)} e^{ipx}. \]
Thus an initial condition where the particles are separated and free is
\[ \psi_{\textrm{\small free}}(x)=c_N\sum_{\s\in\S_N} \prod_{j=1}^N \ph_a(x_{\s(j)};y_j,p_j). \]
where $c_N$ is a normalization constant.

A second choice is for the Bose gas to be initially
confined to a subset of $\R$  and in its ground state.
  Solving (\ref{Schr}) with this initial condition corresponds to the
confined Bose gas freely expanding into all of $\R$.  The ground state
wave function for confinement to $\R^+$ has been computed  by Gaudin \cite{Ga} and confinement
to a hard wall box by Batchelor, \textit{et al.}~\cite{BGOL}.

We now recall a well-known \cite{LL, Su2}  reformulation of the problem.  
Since we seek symmetric solutions, it is sufficient to solve (\ref{Schr})  in the region
\be \CR: -\iy< x_1\le x_2\le \cdots \le x_N<\iy .\label{region}\ee 
In the interior of $\CR$, e.g. $ \CR^o:  x_1<x_2<\cdots <x_N$, 
the $\d$-functions are zero and we have the free Schr\"odinger equation
\be -\sum_i {\pl^2\Psi_\d\ov \pl x_i^2} = i \, {\pl \Psi_\d\ov \pl t}. \label{freeSch}\ee
That is  the effect of the $\d$-functions
are confined to the boundary of $\CR$, e.g.\  on the hyperplanes $x_j=x_{j+1}$, and
their effect can be formulated as a boundary condition on the hyperplanes 
(we use also the fact that $\Psi_\d$ is a symmetric function):

\be\left. \left({\!\!\pl\ov \pl x_{j+1}}-{\pl\ov \pl x_j}\right)\Psi_\d\right|_{x_{j+1}=x_j} = c \left.\Psi_\d\right|_{x_{j+1}=x_j}. 
\label{BC} \ee
Thus the problem is to solve (\ref{freeSch}) in $\CR^o$ subject to the boundary conditions (\ref{BC})
and the initial condition in $\CR$
\be \Psi_\d(x;0) = \prod_{j=1}^N \d(x_j-y_j) \label{IC2}\ee
where $y_j$ satisfy (\ref{yOrder}) and we've  dropped the normalization constant $N!$ since it can be incorporated at the end.
The solution $\Psi_\d$ is given below in (\ref{Psi}).
\section{Bethe Ansatz}
We now explain how the ideas of Bethe Ansatz \cite{Be, LL} can be employed to solve
 the time-dependent problem without using the spectral theory of the operator $H$.  Since we avoid the eigenvalue problem, there
 are no \textit{Bethe equations} in this approach.\footnote{The methods here are an adaptation of the methods of \cite{TW}.}  
 To set notation and to see the argument in its simplest form, we first
 solve $N=2$.  
\subsection{$N=2$}
It is elementary to verify that
\[ \int_{\R}\int_{\R} A(k_1,k_2) e^{i(k_1 x_1+ k_2 x_2)} e^{-i( \ep(k_1)+\ep(k_2)) t} \, dk_1 dk_2 \]
with
\[ \ep(k) = k^2\]
solves (\ref{freeSch}) in $x_1< x_2$.  The insight from Bethe Ansatz  is to add
to this solution another solution with the integration variables permuted,
\[ \int_{\R}\int_{\R} \left[ A_{12}(k_1,k_2) e^{i(k_1 x_1+ k_2 x_2)}
+A_{21}(k_1,k_2) e^{i(k_2 x_1+ k_1 x_2)} \right]e^{-i( \ep(k_1)+\ep(k_2)) t} \, dk_1 dk_2, \]
so that the boundary condition (\ref{BC}) can be applied pointwise to the above integrand. The result is
the condition
\[ ik_2 A_{12} +i k_1 A_{21} - i k_1 A_{12}- ik_2 A_{21} = c(A_{12}+A_{21}), \]
or equivalently,
\[ A_{21}= - {c-i(k_2-k_1)\ov c+i (k_2-k_1)}\, A_{12}. \]
We set
\be S_{\a \b}=S_{\a\b}(k_\a-k_\b)=- {c-i(k_{\a}-k_{\b})\ov c+i (k_{\a}-k_{\b})} \label{Smatrix}\ee
so that the above reads $A_{21}=S_{21} A_{12}$.  It is also convenient to set
\be S(k)=-\frac{c-ik}{c+ik}\, .\label{S}\ee
Observe  that
$S(k)$ extends to a holomorphic function in the lower half-plane and so its
Fourier transform, $\hat S(z)$, is supported in $[0,\iy)$.

With the initial condition (\ref{IC2}) in mind, we choose
\[ A_{12}(k_1,k_2)= e^{-i(k_1 y_1+k_2 y_2)} \]
so the above solution becomes
\be \int_{\R}\int_{\R} \left[ e^{i(k_1 (x_1-y_1)+ k_2 (x_2-y_2))}
+S_{21}(k_2,k_1) e^{i(k_2 (x_1-y_2)+ k_1 (x_2-y_1))} \right]e^{-i( \ep_{k_1}+\ep_{k_2}) t} \, dk_1 dk_2 \label{soln2}\ee
where from now on $dk_j\ra dk_j/2\pi$.

At $t=0$ the first term evaluates to  $\d(x_1-y_1) \d(x_2-y_2)$.
Thus we must show
\[ \int_{\R}\int_{\R}S(k_2-k_1)e^{i k_1(x_2-y_1)+i k_2(x_1-y_2)}\, dk_1 dk_2 =0 \]
for $y_1<y_2$ and $x_1\le x_2$ for the solution (\ref{soln2}) to satisfy the initial condition (\ref{IC2}).
Making the change of variables $k_2\ra k_2+k_1$ and $k_1\ra k_1$ in the above
integral, we have (after performing the resulting integration over $k_1$)
\[ \d(x_1+x_2-y_1-y_2) \, \int_{\R} S(k_2) e^{i k_2(x_1-y_2)}\, dk_2= \d(x_1+x_2-y_1-y_2)  {\hat S}(x_1-y_2).\]
Now ${\hat S}(x_1-y_2)$ is nonzero  only in the region
\[ x_1-y_2\ge 0 \]
and the delta function requires that $x_1+x_2=y_1+y_2$ for a nonzero contribution.
But in $\mathcal{R}$
\[  2 x_1\le x_1+x_2=y_1+y_2< 2 y_2, \]
i.e.\ $x_1<y_2$.  Hence the integral in the region $\mathcal{R}$ is zero;  and thus,  we conclude  $(\ref{soln2})$ is the sought after solution
$\Psi_{\d}(x)$ for $N=2$.
\subsection{General $N$}
Let $\s\in\CS_N$ be a permutation of $\{1, \ldots, N\}$.  Recall that an inversion in a permutation $\s$ is an ordered pair
$\{\s(i),\s(j)\}$ in which $i<j$ and $\s(i)>\s(j)$.
We set
\be A_\s=\prod\left\{ S_{\a \b}: \{\a,\b\}\>\> \textrm{is an inversion in}\>\> \s\right\}\label{Asigma}\ee
where $S_{\a\b}$ is defined by (\ref{Smatrix}).  Thus, for example, $A_{231}=S_{21} S_{31}$ and $A_{\small id}=1$.
We claim the solution that satisfies (\ref{freeSch}) with boundary conditions (\ref{BC}) and initial condition
(\ref{IC2}) is 
\be \Psi_\d(x;t)=\sum_{\s\in\CS_N}\int_{\R}\cdots\int_{\R} A_\s\, \prod_{j=1}^N e^{i k_{\s(j)} (x_j-y_{\s(j)})} \, e^{-it\sum_j \ep(k_j)}\,
dk_1\cdots dk_N. \label{Psi}\ee

First, it is clear that (\ref{Psi}) satisfies (\ref{freeSch}) in $\CR^o$.  As demonstrated  in \cite{LL},  if
$T_i\s$ denotes $\s$ with the entries $\s(i)$ and $\s(i+1)$ interchanged, the boundary conditions will be satisfied provided
that
\[ A_{T_i\s}=S_{\s(i+1)\s(i)} A_\s\]
for all $\s$.  Let us see why this relation holds.  Let $\a=\s(i)$, $\b=\s(i+1)$, and suppose $\a>\b$.  Then $\{\a,\b\}$ is an
inversion for $\s$ but not for $T_i\s$, so $S_{\a\b}$ is a factor in $A_\s$ but not in $A_{{T_i}\s}$, and all other factors
are the same.  Therefore, using $S_{\a\b}S_{\b\a}=1$, we have
\[ A_{{T_i}\s}=S_{\b\a} A_\s =S_{\s(i+1)\s(i)} A_\s.\]
The same identity holds immediately if $\b>\a$, since $\{\b,\a\}$ is an inversion for $T_i\s$ but not for $\s$.

As in the $N=2$ case, the term corresponding to the identity permutation in (\ref{Psi}) satisfies the initial
condition (\ref{IC2}).  Thus to complete the proof we must show
\[\sum_{\s\in\CS_N, \s\neq id}\int_{\R}\cdots\int_{\R} A_\s\, \prod_{j=1}^N  e^{i k_{\s(j)} (x_j-y_{\s(j)})} \,
dk_1\cdots dk_N =0.\]
Let $I(\s)$ denote the integral corresponding to $\s$ term in the above sum.
Recalling the definition (\ref{S}) of $S$, 
 the integrand for $I(\s)$ becomes

\[\prod_{j=1}^N e^{i k_{\s(j)} (x_j-y_{\s(j)})}\;
\prod\left\{ S(k_\a-k_\b): \{\a,\b\}\>\> \textrm{is an inversion in}\>\> \s\right\}.\]
We again use the fact that $S(k)$ extends analytically into the lower half-plane. 

If a number $\g$ appears on the left side of an inversion and never on the right side then in the integrand it appears only in factors of the form $S(k_\g-k_\b)$ and so integrating with respect to $k_\g$ shows that the integrand is zero unless 
\be x_{\s\inv(\g)}\ge y_\g.\label{left}\ee 
Similarly, if a number $\d$ appears on the right side of an inversion and never on the left side then in the integrand it appears only in factors of the form $S(k_\a-k_\d)$ and so integrating with respect to $k_\d$ shows that the resulting integrand is zero unless 
\be x_{\s\inv(\d)}\le y_\d.\label{right}\ee 

We show $I(\s)=0$ by induction on $N$. If $\s(1)=1$ then 1 does not appear in an inversion, we can integrate with respect to $k_1$, and we are reduced to the case $N-1$. So assume $\s(1)=\g>1$. Then $(\g,\,1)$ is an inversion and there is no inversion of the form $(\a,\,\g)$ because $\g$ appears in slot 1. So we can apply (\ref{left}) with this $\g$ and get the resulting integrand is zero unless
\[x_1\ge y_\g.\]

Next we observe as before that $(\g,\,1)$ is an inversion, but now that 1, which appears on the right side of the inversion, cannot appear on the left side of an inversion (obviously). So we can apply (\ref{right}) with $\d=1$ and get
\[x_{\s\inv(1)}\le y_1.\] 
Since $y_1<y_\g$ the two inequalities give $x_{\s\inv(1)}<x_1$, which cannot happen.
This completes the proof of (\ref{Psi}).

The limit of impenetrable bosons
is  the limit $c\ra\iy$.  In this limit $S_{\a\b}\ra -1$ and
$A_\s\ra(-1)^{ inv(\s)}$ where $\textrm{inv}(\s)$ is the number of inversions in $\s$.  Thus in $\mathcal{R}$
\be \Psi_\d(x;t)\ra\Psi_\iy(x;t)=\int_\R\cdots \int_\R \det\left(e^{i k_\a(x_\b-y_\a)-it\ep_{k_\a}}  \right)\, dk_1\cdots dk_N \label{impenetrablePsi}\ee
and the extension  to  $\R^N$ follows by requiring $\Psi_\iy$ to be a symmetric function of $x_j$.
\newpage
 \noindent{\textbf{Acknowledgements:}  This work was supported by the National Science Foundation under grants
 DMS--0553379 (first author) and DMS--0552388 (second author).


\begin{thebibliography}{99}

\bibitem{BGOL} Batchelor, M.T., Guan, X.W., Oelkers, N., Lee, C.: The $1$D interacting Bose
gas in a hard wall box. J.\ Phys.\ A.: Math.\ Gen.\ \textbf{38}, 7787--7806, 2005.


\bibitem{Be} Bethe, H.A.: On the theory of metals, I. Eigenvalues and eigenfunctions of a linear
chain of atoms (German). Zeits.\ Phys.\ \textbf{1931}, 205--226 (1931).
[English translation appears in Bethe, H.A.: \textit{Selected Works of Hans A.~Bethe With Commentary},
World Scientific, Singapore, 1996, pgs.~689--716.]

\bibitem{BDZ} Bloch, I., Dalibard, J., Zwerger, W.: Many-body physics with ultracold gases.
Rev.\ Mod.\ Phys.\ \textbf{80}, 885--964 (2008).

\bibitem{Ga} Gaudin, M.:  Boundary energy of a Bose gas in one dimension.
Phys.\ Rev.\ A \textbf{4}, 386--394 (1971).

\bibitem{KBI} Korepin, V.E., Bogoliubov, N.M., Izergin, A.G.: \textit{Quantum Inverse Scattering Method and Correlation Functions}, Cambridge University Press, 1993.

 \bibitem{L} Lieb, E.H.: Exact analysis of an interacting Bose gas. II. The excitation spectrum.
 Phys.\ Rev.\ \textbf{130}, 1616--1624 (1963).

\bibitem{LL} Lieb, E.H., Liniger, W.: Exact analysis of an interacting Bose gas. I.
 The general solution and the ground state. Phys.\ Rev.\ \textbf{130}, 1605--1616 (1963).
 
 
 \bibitem{Se} Seiringer, R., Yin, J.: The Lieb-Liniger model as a limit of dilute bosons in
 three dimensions.  Commun.\ Math.\ Phys.\ DOI: 10.1007/s00220-008-0521-6.
 
 \bibitem{Su1} Sutherland, B.: Further results for the many-body problem in one dimension.  Phys.\ Rev.\ Lett.\
 \textbf{20}, 98--100 (1968).
 
 \bibitem{Su2} Sutherland, B.: \textit{Beautiful Models: 70 Years of Exactly Solved Quantum Many-Body
 Problems}, World Scientific Publishing Co., 2004.
 
\bibitem{TW} Tracy, C.A., Widom, H.: Integral formulas for the asymmetric simple exclusion process.
Commun.\ Math.\ Phys.\ \textbf{132}, 815--844 (2008).

\bibitem{Ya} Yang, C.N.: Some exact results for the many-body problem in one dimension with replusive
delta-function interaction. Phys.\ Rev.\ Lett.\ \textbf{19}, 1312--1315 (1967).

\bibitem{YY} Yang, C.N., Yang, C.P.: Thermodynamics of a one-dimensional system of bosons with replusive
delta-function interaction. J.\ Math.\ Phys.\ \textbf{10}, 1115--1122 (1969).

\end{thebibliography}
\end{document}